\documentclass[12pt]{article}
\usepackage{epsfig}
\begin{document}

\centerline {\Large \bf Corrugation of Roads}
\vskip 1.0 true cm
\centerline{ \bf Joseph A. Both and Daniel C. Hong }
\vskip 0.2 true cm
\centerline{ Physics, Lewis Laboratory,
Lehigh University, Bethlehem, Pennsylvania 18015}
\vskip 0.3 true cm
\centerline{ {\bf Douglas A. Kurtze}, kurtze@golem.phys.ndsu.nodak.edu}
\vskip 0.2 true cm
\centerline{ Department of Physics, North Dakota State University,
Fargo, North Dakota 58015}
\vskip 0.3 true cm

\begin{abstract}

We present a one dimensional model for the development of corrugations in
roads subjected to compressive forces from a flux of cars.  The cars are
modeled as damped harmonic oscillators translating with constant horizontal
velocity across the surface, and the road surface is subject to diffusive
relaxation.  We derive dimensionless coupled equations of motion for the
positions of the cars and the road surface $H(x,t)$, which contain two
phenomenological variables: an effective diffusion constant $\Delta(H)$
that characterizes the relaxation of the road surface, and a function
$a(H)$ that characterizes the plasticity or erodibility of the road bed.
Linear stability analysis shows that corrugations grow if the speed of the
cars exceeds a critical value, which decreases if the flux of cars is
increased.  Modifying the model to enforce the simple fact that the normal
force exerted by the road can never be negative seems to lead to
restabilized, quasi-steady road shapes, in which the corrugation amplitude
and phase velocity remain fixed.
\end{abstract}

\section{Introduction}


It is commonly observed that under the influence of a flow of traffic,
dirt roads develop regular corrugations of longitudinal ``pitch,'' or
wavelength, between 0.5 and 1 m, and amplitude up to 50 mm \cite{SSC82}.
At first glance such an instability of the road surface might seem
counterintuitive, as one might guess that a flow of traffic would exert
downward forces on the surface which tend to smooth and compact the road
bed, thereby suppressing pattern development rather than promoting it.  Yet
irregular, rough roads do not in fact heal themselves, but instead become
progressively rougher and more corrugated under a flow of cars.  Similar
phenomena also occur, though sometimes on different length and time scales,
on paved roads, on railroad rails \cite{Mather}, and on the rollers used to
calendar paper \cite{Papad}.  The purpose of this paper is to investigate
this phenomenon using a simple, tractable physical model.

An early attempt at explaining road corrugation is due to Relton
\cite{Relton}, who proposed that the underlying instability mechanism is a
``relaxation oscillation,'' caused essentially by stick-slip dynamics.
According to this view, a moving wheel pushes grains ahead of it.  The
grains pile up in front of the wheel and form a heap.  When the heap grows
large enough, the wheel sticks momentarily, and then slips, running over the
heap and leaving it behind as a ridge.  For a given uniform speed, this
stick-slip process will be fairly periodic, and so will generate equidistant
ridges.

A different picture was provided by Mather \cite{Mather}, who argued that
the origin of the road instability is the bouncing motion of the wheel,
caused by random irregularities on the ground.  When the bouncing occurs,
the car is projected upward along a certain angle and is airborne for a
brief time.  When it then strikes the ground, the car creates a crater and
the motion then repeats itself.  According to this picture, it is not the
piling up of grains ahead of the vehicle that is responsible for the
instability, but the impact stress of the vehicle on the ground.  The
wavelength of the resulting corrugations will be determined by the
competition between the typical distance the car flies over the ground and
the size of the crater generated by the impact stress, which should in turn
depend on the hardness of the ground and the relaxation time of the ejected
grains.  Mather's picture is similar to other surface instabilities
involving granular materials, in particular the ripple patterns in wind
blown sand \cite{Bagnold,PT90,NO93,KBH00}, where ejected grains are
carried away
by the wind and land in a place far from the ejection point.  In such a
nonlocal model, what sets the wavelength of the ripple is the ratio of the
flux of the grains to the appropriately scaled saltation length, i.e., the
distance that an ejected grain is carried by the wind.

The model we present in this paper builds on Mather's picture, but we
ignore any nonlocal transport of grains along the road, and we assume that
the cars, and more specifically their wheels, generally do not lose contact
with the road.  Instead, we model the cars simply as masses attached to
damped springs.  We assume that the downward contact forces exerted on the
road by the wheels causes a permanent downward deflection of the road
surface, by either erosion or plastic deformation.  In either case we model
the effect of the contact force on the road as a simple proportionality
between the deformation at any point on the road and the force exerted on
that point, with a phenomenological ``softness'' parameter as the
proportionality constant.  We also include a diffusive relaxation that tends
to smooth the road surface, which may come about, for instance, as a result
of rain.  We find that the diffusive relaxation is a stabilizing effect, as
one would expect, but that its presence or absence generally has no
qualitative effect on the corrugation phenomena.  What is important,
however,
is the phenomenon of hardening.  We expect that the softness parameter, and
also the diffusion coefficient if it is present, will decrease as the road
compacts, so that the road becomes less susceptible to the passage of more
cars once the first several have compacted it and perhaps produced
corrugations.  This turns out to be a stabilizing effect, narrowing the
parameter range in which corrugations occur.

Since we are modeling the cars, which in reality are complicated mechanical
systems, by simple damped harmonic oscillators, the question quickly arises
as to what the appropriate natural frequency might be.  Given the observed
pitch of road corrugations of 0.5 to 1 m, and the presumed speeds in the
range of, say, 10 to 25 m/s of the cars that produce them, we expect that
the important oscillation modes must have natural frequencies of about
0.02 to 0.1 s.  This is one or two orders of magnitude shorter than the
frequency of the car body bouncing on its suspension, so that is unlikely
to be the relevant oscillation; indeed drivers normally adjust to the
roughness of the road they are on and slow down to avoid such bouncing.
This suggests that the relevant mode may be that of the wheel attached to
the suspension.  Alternatively, the important oscillation may be an
elastic deformation of the wheel itself \cite{CJ71}.  This could account for
an observed difference in the pitch of corrugations produced by hard and
soft tires \cite{SSC82}.

\section{Model}

\begin{figure}[tb]
\begin{center}
\epsfig{file=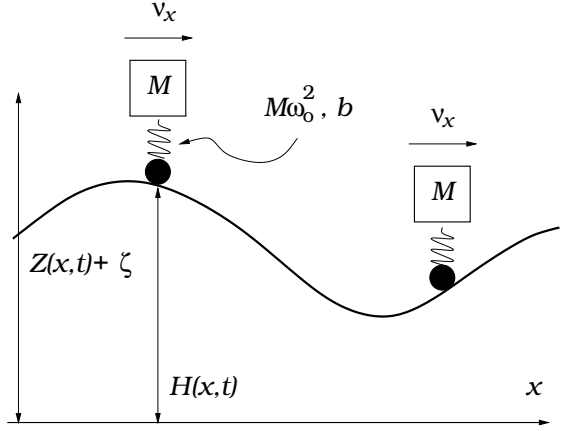,height=6cm,angle=0.0,clip=}
\end{center}
\caption{A schematic picture of the corrugation of roads.  The road surface
is subjected to a flux of cars with horizontal speed $v_x$, and the cars are
modeled as damped harmonic oscillators with mass $M$, spring constant
$M \omega_0^2$, and damping coefficient $b$.  The heights of the cars and
the surface are $Z(x,t) + \zeta$ and $H(x,t)$ respectively, where $\zeta$ is
the equilibrium length of the spring.}
\label{fig1}
\end{figure}

Imagine a flux of cars traveling with some average horizontal velocity
$v_x$ along a road surface whose height above some arbitrary zero level is
given as a function of time $t$ and position $x$ along the road by $H(x,t)$;
see Fig. \ref{fig1}.  The cars are supported on springs, such that the
natural
angular frequency of vertical oscillation of the cars is $\omega_0$.
Further, we assume that the springs are damped with damping constant $b$.
Let $Z(x,t)$ be the height of a moving car, relative to a zero level chosen
such that $H(x,t) - Z(x,t)$ is the amount by which the springs are
compressed.  In a frame of reference moving horizontally with the car, the
vertical component of the car's equation of motion comes simply from
Newton's second law,
\begin{eqnarray}
  M \frac{d^2}{dt^2} Z(x,t) + b \frac{d}{dt}[Z(x,t)-H(x,t)]
  = \nonumber \\
  M \omega_0^2 [H(x,t)-Z(x,t)].
\end{eqnarray}

The time derivatives here are total derivatives; that is, the relevant
value of $x$ is time-dependent, since this equation follows the vertical
motion of a single car.  We may convert this into an equation for the
height of the car at a fixed $x$ by replacing the total time derivative
$d/dt$ with $\partial/\partial t + v_x (\partial/\partial x)$.  Thus
in a reference frame that is fixed to the road bed, the vertical equation
of motion is
\begin{eqnarray}
  M \left(\frac{\partial}{\partial t} + v_x\frac{\partial}{\partial
x}\right)^2 Z
  + b \left(\frac{\partial}{\partial t} + v_x\frac{\partial}{\partial
x}\right) (Z - H) \nonumber \\
  + M \omega_0^2 (Z - H) = 0.
\end{eqnarray}
Note that we have neglected any possible variation in the horizontal
velocity $v_x$ of the cars.

We must now write an evolution equation for the height $H(x,t)$ of the road
surface.  We assume first that the road surface sinks at a rate which is
proportional to the downward force on the road.  This may be due to
compaction of the road bed under the surface or to ejection of loose grains
at the surface as a car passes; in either case we expect that the
proportionality constant between the downward force on the road and its
rate of sinking will decrease as the road sinks.  That is, the road should
harden as more and more cars pass over it.  In addition, we assume that
there is a ``diffusive'' relaxation process which tends to even out any
roughness in the road surface.  This can result from the action of wind
or rain, and may also contain a contribution from the passage of the cars,
as the loosely connected grains at the surface are fluidized by the flow
of cars.  Again, we expect the effective diffusion coefficient to decrease
as cars pass and the road hardens.  With these assumptions, the equation
of motion for the road height reads
\begin{equation}
  \frac{\partial H}{\partial t} = D(H) \frac{\partial^2 H}{\partial x^2}
   - a(H) \left[ M g + M \omega_0^2 (H-Z) \right],
\end{equation}
where $Mg$ is the weight of a car.  We are neglecting the geometrical
distinction between $\partial H/\partial t$ and the rate of motion of the
road surface normal to itself, and also the fact that the compressive force
is not really vertical when $H$ is not constant.  These are satisfactory
approximations provided the vertical scale of the surface corrugations is
much smaller than the horizontal scale.  However, these effects should be
included in any model of road corrugation which is nonlinear in the
amplitude of the corrugation.

\begin{figure}[tb]
\begin{center}
\epsfig{file=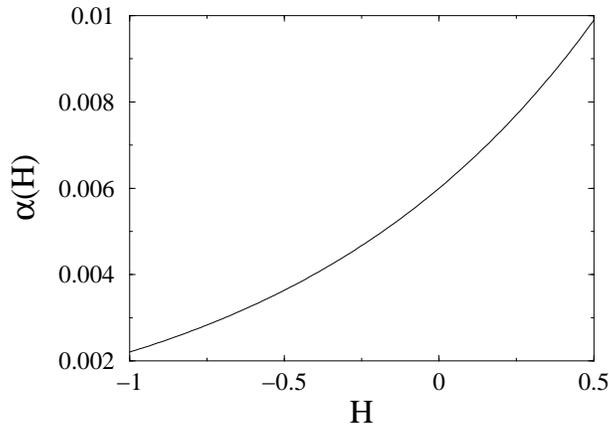,height=6cm,angle=0.0,clip=}
\end{center}
\caption{A physically plausible form for the softness or compactivity
function $a(H)$.  We require $a$ to be an increasing function of $H$ which
either approaches zero for $H \to -\infty$ or vanishes for $H$ below some
fixed, finite limit.}
\label{fig2}
\end{figure}

The proportionality factor $a(H)$ above represents the softness of the
road, that is, its responsiveness to compressing forces.  It should
include the flux of cars as a multiplicative factor, as the rate of road
sinking should also depend on the density of the traffic it sustains.
The compression term should be replaced by zero whenever the quantity in
brackets is negative, since that represents the situation in which the
cars are airborne, so that the compressive force on the road would really
be zero rather than negative.  Requiring that the road harden as it compacts
means that $a$ (and probably also $D$) should decrease as $H$ decreases, so
we want $a(H)$ to be positive and an increasing function of $H$.
A physically acceptable form of $a(H)$ is shown in Fig. \ref{fig2}.

Before proceeding, we first nondimensionalize the equations of motion for
$Z$ and $H$.  We choose the time scale to be $1/\omega_0$, the horizontal
length scale to be $v_x/\omega_0$, and the vertical length scale to be
$g/\omega_0^2$.  The equation of motion for the cars then becomes
\begin{equation}
 \left( \frac{\partial}{\partial t} + \frac{\partial}{\partial x} \right)^2
Z
  + 2 \Gamma  \left( \frac{\partial}{\partial t} + \frac{\partial}{\partial
x} \right) (Z-H)
  + Z - H = 0,
\label{bigzeqn}
\end{equation}
and the equation for the road surface is
\begin{equation}
  \frac{\partial H}{\partial t} = \Delta(H) \frac{\partial^2 H}{\partial
x^2}
    - \alpha(H) (1 + H - Z),
\label{bigheqn}
\end{equation}
where the new dimensionless parameters are given by
\begin{eqnarray}
  \Gamma = b / 2 M \omega_0, \qquad \Delta(H) = (\omega_0 / v_x^2) \, D(H),
  \nonumber \\
   \qquad \alpha(H) = M \, \omega_0 \, a(H).
\label{parmdefs}
\end{eqnarray}
The parameter $\Gamma$ is a property of the cars alone; the cars' springs
are underdamped for $\Gamma<1$.  Since the instability is controlled by
the interplay of essentially two mechanisms, ejection of grains from the
surface or compaction of the road by the flux of cars and subsequent
relaxation of the road surface due to diffusion, $\Gamma$ and $\Delta$ will
control the dynamics of the surface.  As we will see, the hardening of the
road also plays an important role in the development of the instability.

\section{Linear Stability Analysis}

We may solve the system of partial differential equations (\ref{bigzeqn})
and (\ref{bigheqn}) numerically, and we will discuss this below, but useful
information regarding the behavior of the system may also be extracted from
an approximate linear stability analysis.  The first step in the analysis
is to recognize that as time increases, the spatially averaged values of
$Z(x,t)$ and $H(x,t)$ will tend to decrease, reflecting the gradual settling
of the road bed.  We denote this spatially averaged, i.e. spatially
independent, {\it quiescent solution\/} by $H_0(t)$ and $Z_0(t)$.
Substituting this solution into the equations of motion, we find that it
satisfies
\begin{equation}
  \ddot Z_0 + 2 \Gamma (\dot Z_0 - \dot H_0) + Z_0 - H_0 = 0.
\label{Z0eqn}
\end{equation}
and
\begin{equation}
  \dot H_0 = - \alpha(H_0) (1 + H_0 - Z_0),
\label{H0eqn}
\end{equation}
The full solution may now be written as
\begin{equation}
  H(x,t) = H_0(t) + h(x,t), \qquad Z(x,t) = Z_0(t) + z(x,t),
\end{equation}
in which the functions $h(x,t)$ and $z(x,t)$ will carry information about
pattern development.  Substituting these forms of $H(x,t)$ and $Z(x,t)$ into
(\ref{bigzeqn}) and (\ref{bigheqn}), using (\ref{Z0eqn}) and (\ref{H0eqn})
for the time evolution of $Z_0$ and $H_0$, and expanding to first order in
$h$ and $z$, we obtain the following linearized equations of motion:
\begin{equation}
  \left( \frac{\partial}{\partial t} + \frac{\partial}{\partial x}
\right)^2 z
  + 2 \Gamma \left( \frac{\partial}{\partial t} + \frac{\partial}{\partial
x} \right) (z-h)
  + z - h = 0,
\label{zeqn}
\end{equation}
\begin{equation}
  \frac{\partial h}{\partial t} = \Delta \frac{\partial^2 h}{\partial x^2}
    - \alpha (h - z) - \beta h,
\label{heqn}
\end{equation}
where $\alpha$ and $\Delta$ are evaluated at $H = H_0$, and $\beta$ is
given by
\begin{equation}
  \beta = (1 + H_0 - Z_0) \, \alpha'(H_0)
        = - \frac{d \hbox{ln} \alpha(H_0)}{dt}.
\label{betadef}
\end{equation}
The last equality here follows from (\ref{H0eqn}).  Note that $\beta$
should be positive, since we expect that $\alpha$ will increase with $H$ and
decrease with time.

If $\alpha$ and $\Delta$ are nontrivial functions of $H$, then it is not
simple to solve the linearized equations (\ref{zeqn}) and (\ref{heqn}),
because the coefficients in the latter equation are functions of time.
However, we may perform an {\it approximate\/} stability analysis by
regarding $\alpha$, $\beta$, and $\Delta$ as constants, at least for
short time intervals, and calculating the linear growth rates that
obtain when those parameters have their current values.  Thus we will
determine {\it time dependent\/} growth rates and phase velocities,
provided we can ascertain the time dependent forms of $\alpha$ and $\Delta$.

We proceed by assuming both $h$ and $z$ to be proportional to
$\exp(ikx + \sigma t)$, where the linear growth rate $\sigma$ is in general
complex.  As usual, $Re[\sigma]$ is the exponential growth or decay rate of
the amplitude of a perturbation with wave number $k$, and $Im[\sigma]$ is
related to the phase velocity $c$ of that mode through $c=-Im[\sigma]/k$.
This substitution yields
\begin{equation}
  [(\sigma + i k)^2 + 2 \Gamma (\sigma + i k) + 1] z
    = [1 + 2 \Gamma (\sigma + i k)] \, h
\end{equation}
and
\begin{equation}
  \sigma h = - \Delta k^2 h - \alpha (h - z) - \beta h.
\end{equation}
Eliminating $h$ between these equations gives the stability relation
\begin{equation}
  [(\sigma + i k)^2 + 2 \Gamma (\sigma + i k) + 1] \,
  (\beta + \sigma + \Delta k^2) + \alpha \, (\sigma + i k)^2 = 0,
\label{stabrel}
\end{equation}
from which we can find the growth rates and phase velocities of spatially
sinusoidal perturbations in terms of their wave numbers.

It is possible to determine the stability boundary for the model, at least
parametrically, from (\ref{stabrel}).  That is, we could determine the locus
in parameter space on which the real part of the linear growth rate
$\sigma$, as a function of $k$, has a global maximum at height
$Re[\sigma]=0$.  However, it is much more instructive to simplify the
already approximate problem further by taking $\alpha$ and $\Delta$ to be
small, as we expect to be the case once the road has had a chance to harden
sufficiently.  To set the stage for this calculation, imagine setting
$\alpha=0$ in (\ref{stabrel}).  The cubic equation for $\sigma$ would then
factor.  Two of the solutions would always have negative real parts, since
$\sigma + i k$ for these two solutions would be a root of a quadratic with
positive coefficients; these represent decaying perturbation modes.  The
third would be $\sigma = -\beta - \Delta k^2$, so this mode too would always
be stable unless $\beta$ is also small, on the same order as $\alpha$ or
smaller.  Note that small $\alpha$ does {\it not\/} necessarily imply that
$\beta$ must be small, because $\beta$ is the {\it logarithmic\/} derivative
of $\alpha$.

Since we are interested in finding parameter ranges for which corrugations
grow, we now focus on the case where $\alpha$, $\beta$, and $\Delta$ are all
small.  As we just saw, only one of the three solutions for $\sigma$ can
possibly be positive; to leading order in the small parameters it is given
by
\begin{equation}
  \sigma = -\beta - \Delta k^2 + \alpha k^2
     \frac{(1-k^2) - 2 i \Gamma k}{(1-k^2)^2 + 4 \Gamma^2 k^2}.
\label{smallparms}
\end{equation}

\begin{figure}[tb]
\begin{center}
\epsfig{file=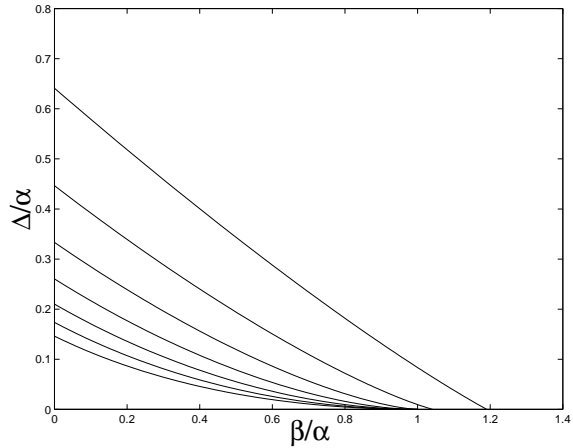,height=6cm,angle=0.0,clip=}
\end{center}
\caption{ The approximate stability diagram in the
$(\beta/\alpha)$--$(\Delta/\alpha)$ plane, obtained from (\ref{smallparms}),
for values of the damping parameter $\Gamma$, ranging from 0.3 (top) to 0.9
(bottom) in steps of 0.1.  All stability boundaries For $\Gamma \ge 0.5$
pass
through the point (0,1).  The flat road surface is stable above and to the
right of the boundary curve for the appropriate $\Gamma$ value.}
\label{fig3}
\end{figure}

We may locate an approximate stability boundary by seeking combinations of
parameters for which the real part of this $\sigma$ has a maximum, as a
function of wave number $k$, at $Re[\sigma]=0$.  Thus we solve the equations
$Re[\sigma] = 0$ and $d(Re[\sigma])/dk = 0$ to obtain the critical values of
$\beta/\alpha$ and $\Delta/\alpha$ parametrically in terms of $k$.  The
results are shown for several values of the damping parameter $\Gamma$ in
Fig. \ref{fig3}.  As is clear from (\ref{smallparms}), both $\beta$ and
$\Delta$ represent stabilizing effects, so for a given $\Gamma$, the flat
road surface is stable above and to the right of the stability boundary
curve for that $\Gamma$.  The stability boundaries move down and left as
$\Gamma$ is increased, so damping of the car's springs also tends to
stabilize the flat road surface.  One can show explicitly that the stability
boundary reaches $\Delta = 0$ at $\beta/\alpha = [4\Gamma(1+\Gamma)]^{-1}$,
and it reaches $\beta = 0$ at $\Delta/\alpha = [4\Gamma(1-\Gamma)]^{-1}$ for
$\Gamma<1/2$ or $\Delta/\alpha = 1$ for $\Gamma>1/2$.  Moreover, for small
$\Gamma$ -- that is, when the oscillations of the car springs are very
slightly damped -- the stability boundary is given approximately by the line
$\Delta + \beta = \alpha/4\Gamma$.

To find the wave numbers of the most important perturbations, we first
note from (\ref{smallparms}) that the parameter $\beta$ only enters
the expression for the growth rate additively.  As a result, the wave
number $k_{max}$ at which the real part of $\sigma$ has its maximum is
independent of $\beta$; it is given explicitly by
\begin{equation}
  \frac{\Delta}{\alpha} =
     \frac{(1-k_{max}^2)^2 - 4 \Gamma^2 k_{max}^4}
          {[(1-k_{max}^2)^2 + 4 \Gamma^2 k_{max}^2]^2}.
\label{kmaxeqn}
\end{equation}

\begin{figure}[tb]
\begin{center}
\epsfig{file=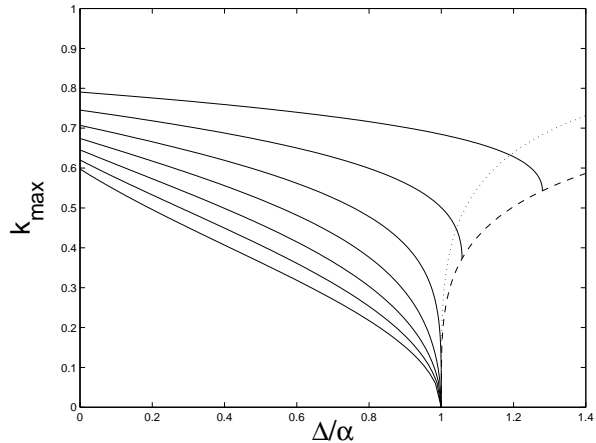,height=6cm,angle=0.0,clip=}
\end{center}
\caption{The wave number of the perturbation whose linear growth rate is
maximal, plotted as a function of $\Delta/\alpha$ for various values of the
damping parameter $\Gamma$ ranging from 0.3 (top) to 0.9 (bottom) in steps
of 0.1.  The dashed curve marks where the local maximum in $Re[\sigma(k)]$
merges with a local minimum and vanishes.  Between it and the dotted curve,
$Re[\sigma]$ has a local maximum at the indicated $k$, but the {\it
global\/}
maximum is at $k=0$.}
\label{fig4}
\end{figure}

Fig. \ref{fig4} shows $k_{max}$ as a function of $\Delta/\alpha$ for several
values of the damping constant $\Gamma$.  (Below and to the right of the
dotted curve, the maximum growth rate is negative for any positive value of
$\beta$, so the flat road surface can only be unstable when $\Delta/alpha$
is to the left of the dotted curve.)  We see
that the wave number of the most rapidly growing mode decreases with
increasing damping $\Gamma$; it also decreases with increasing
$\Delta/\alpha$, although the dependence on $\Delta/\alpha$ is weak when
$\Gamma$ is small.  In fact, for small $\Gamma$ we always have
$k_{max} = 1 - \Gamma + O(\Gamma^2)$.  In no case do we find $k_{max}$
to be above 1.  Recall that $k$ is the wave number of the perturbation
in units of $\omega_0/v_x$.  Thus the fact that $k_{max}$ is always less
than 1 means that the wavelength of the most rapidly growing perturbation
is always greater than $(2\pi/\omega_0)\,v_x$, the distance a car travels
in one period of its natural oscillation.

From the imaginary part of $\sigma$ we obtain the drift velocity $c$ of a
perturbation of wave number $k$ (in units of $v_x$),
\begin{equation}
  c = -Im[\sigma] / k
    = \frac{2 \alpha \Gamma k^2}{(1-k^2)^2 + 4 \Gamma^2 k^2}.
\end{equation}
Note the factor $\alpha$:  since $\alpha$ is small, the drift velocity is
small compared to $v_x$, the speed of the cars.  Also, since $\alpha$
contains a factor of the number of cars passing per unit time, the drift is
actually a fixed distance per passing car, rather than per unit time.  The
drift velocity is positive, so that corrugations are pushed in the direction
that traffic is moving as they develop.  It reaches its maximum value of
$\alpha/2\Gamma$ at $k = 1$.

\section{Numerical Results}

Since the linear stability analysis of the preceding section is only
approximate, we have found it useful to check its results by carrying out
numerical simulations of the full model given by Eqns. (\ref{bigzeqn}) and
(\ref{bigheqn}).  We choose a wave number $k$, choose a sinusoidal form
for the initial $H(x)$ and $Z(x)$ with small amplitudes, and then integrate
the equations on an interval of length $2\pi/k$ with periodic boundary
conditions.  We the monitor the amplitudes and phase velocities of $H$ and
$Z$ as time passes.  The logarithmic derivative of the amplitudes with
respect to time gives the instantaneous exponential growth rate $\sigma(t)$.

For the non-constant $\alpha$ cases we use the simple ansatz
\begin{equation}
  \alpha(H) = \alpha_0 \, \exp(\epsilon H), \qquad \epsilon > 0,
\label{alphaeqn}
\end{equation}
which has the qualitative form shown in Fig. \ref{fig2} and is
mathematically tractable.  The parameter $\epsilon$ controls the rate at
which the soil hardens.  We may solve for the time evolution of the
quiescent road level $H_0$ provided $\alpha$ is small, for then $Z_0$ varies
on a much shorter time scale than $H_0$, and so $Z_0$ should always remain
near $H_0$.  Then (\ref{H0eqn}) reduces to
\begin{equation}
  \dot H_0 = - \alpha(H_0) = - \alpha_0 \, \exp(\epsilon H_0).
\end{equation}
This can be integrated immediately to yield
$\exp(-\epsilon H_0) = \epsilon \alpha_0 (t + t_0)$, where $t_0$ is a
constant of integration.  From this we get
\begin{equation}
  \alpha = \frac{1}{\epsilon (t + t_0)},
\end{equation}
and from the definition (\ref{betadef}) of $\beta$ we then get
\begin{equation}
  \beta = \frac{1}{t + t_0} = \epsilon \, \alpha,
\end{equation}
so that the combination $\beta/\alpha$ that appears in the approximate
linear stability analysis is just the constant $\epsilon$.

\begin{figure}[tb]
\begin{center}
\epsfig{file=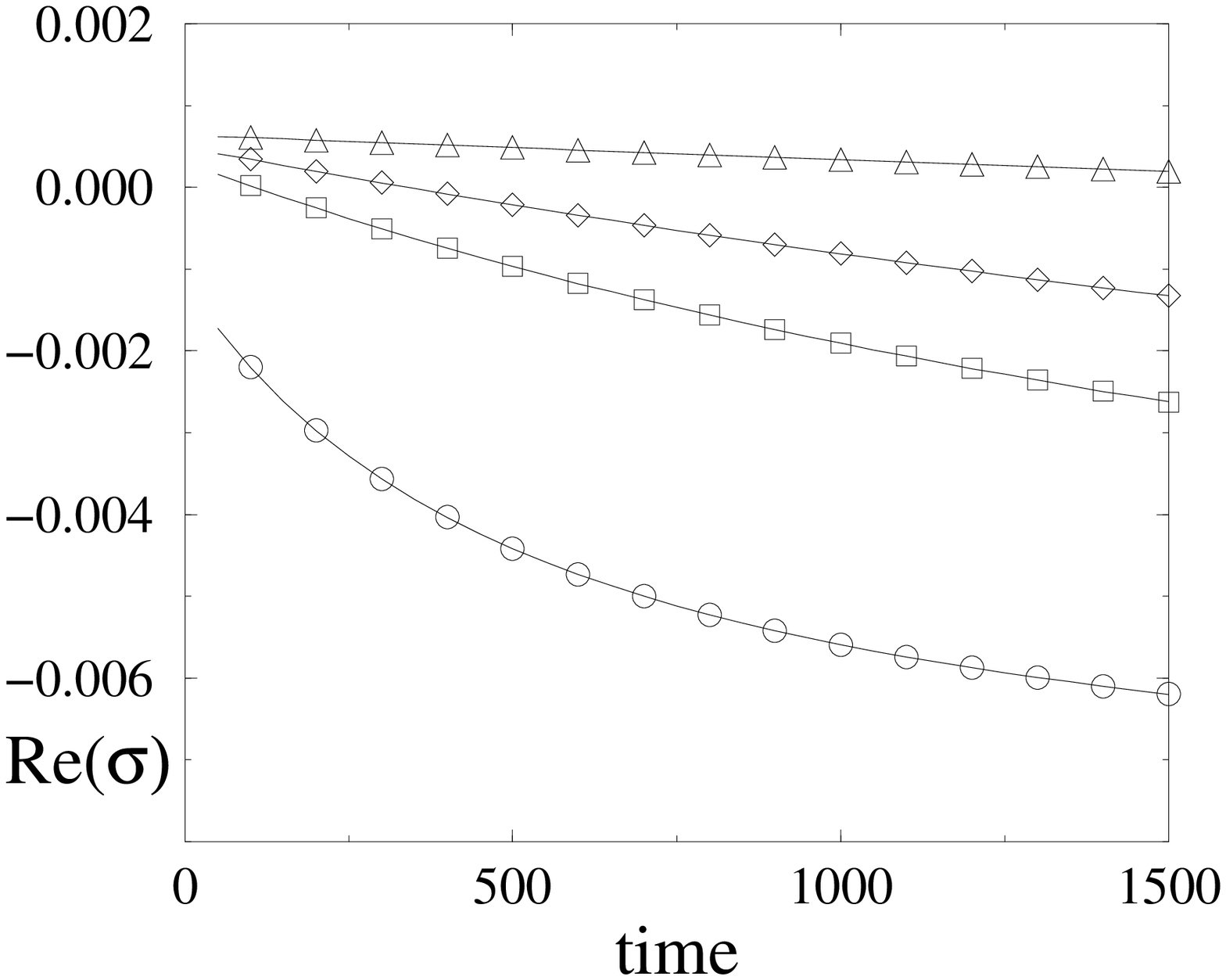,height=6cm,angle=0.,clip=}
\epsfig{file=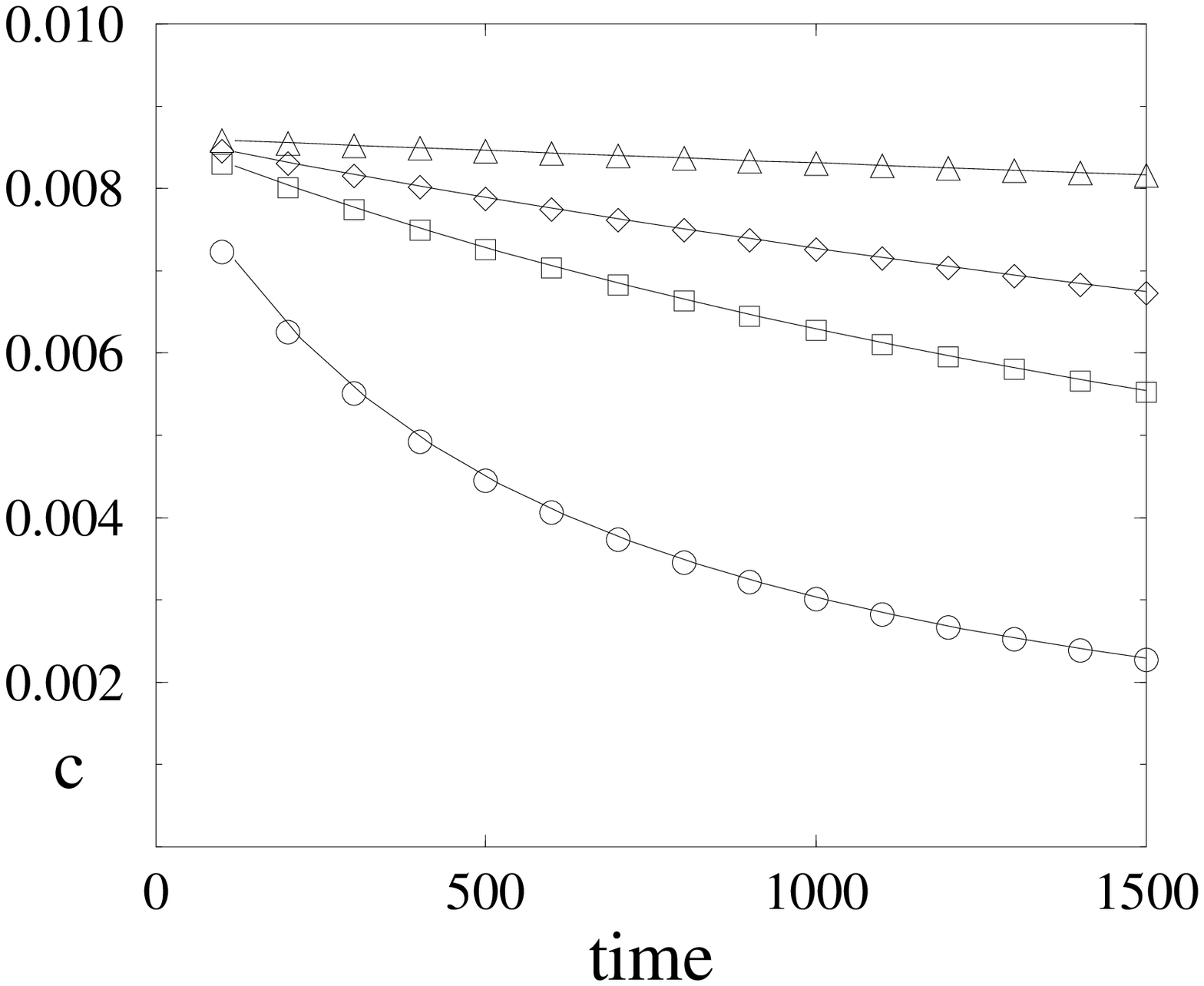,height=6cm,angle=0.,clip=}
\end{center}
\caption  {(a) Linear growth rate $Re[\sigma]$ and (b) phase velocity $c$
vs. time for single mode solutions in the case
$\alpha = \alpha_0 \exp(\epsilon H)$ with $k = 0.90$, $\alpha_0 = 0.004$,
$\Gamma = 0.1$, $\Delta = 0.01$, $\epsilon = 0.5$ circle, $0.1$ square,
$0.05$ diamond, $0.01$ triangle.  Continuous lines are from numerical
solution of the full equations, symbols from the approximate linear
stability analysis.}
\label{fig5}
\end{figure}

Figs. \ref{fig5}a and \ref{fig5}b show typical numerical and theoretical
determinations of the growth rate $Re[\sigma(t)]$ and phase velocity
$c(t) = -Im[\sigma(t)]/k$ for $\alpha$ given by (\ref{alphaeqn}) above
and $\Delta$ constant, for various values of the hardening parameter
$\epsilon = \beta/\alpha$.  As expected, for a given value of $k$ the
agreement is quite strong; this continues to hold over a range of $\alpha_0$
and $\epsilon$.

It is clear from the the linear stability analysis that if the diffusion
parameter $\Delta$ remains finite as $\alpha$ decreases to zero, or more
generally if $\alpha/\Delta$ goes to zero as the road compacts, then we will
eventually reach a situation in which diffusion dominates the dynamics.
From that point on, any corrugation in the road will decay.  On the other
hand, we certainly expect that as the road compacts and its surface hardens,
the hardening should inhibit the lateral transfer of material which we have
modeled as diffusion, as well as further compaction of the road.  We see,
then, that in order to get nontrivial corrugations on the road surface, we
must have $\Delta$ decrease at least as rapidly as $\alpha$ as the road
compacts.

\begin{figure}[tb]
\begin{center}
\epsfig{file=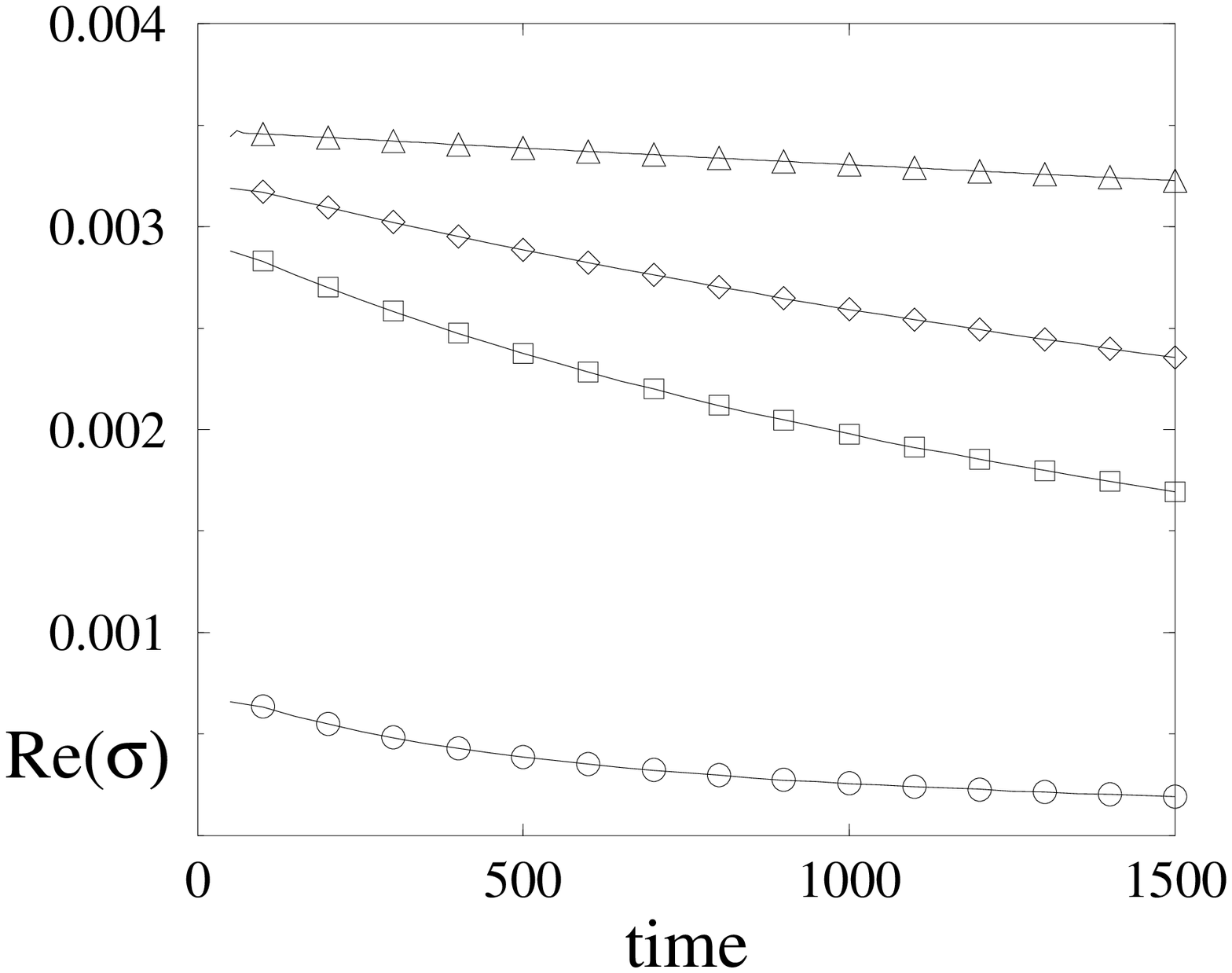,height=6cm,angle=0.,clip=}
\epsfig{file=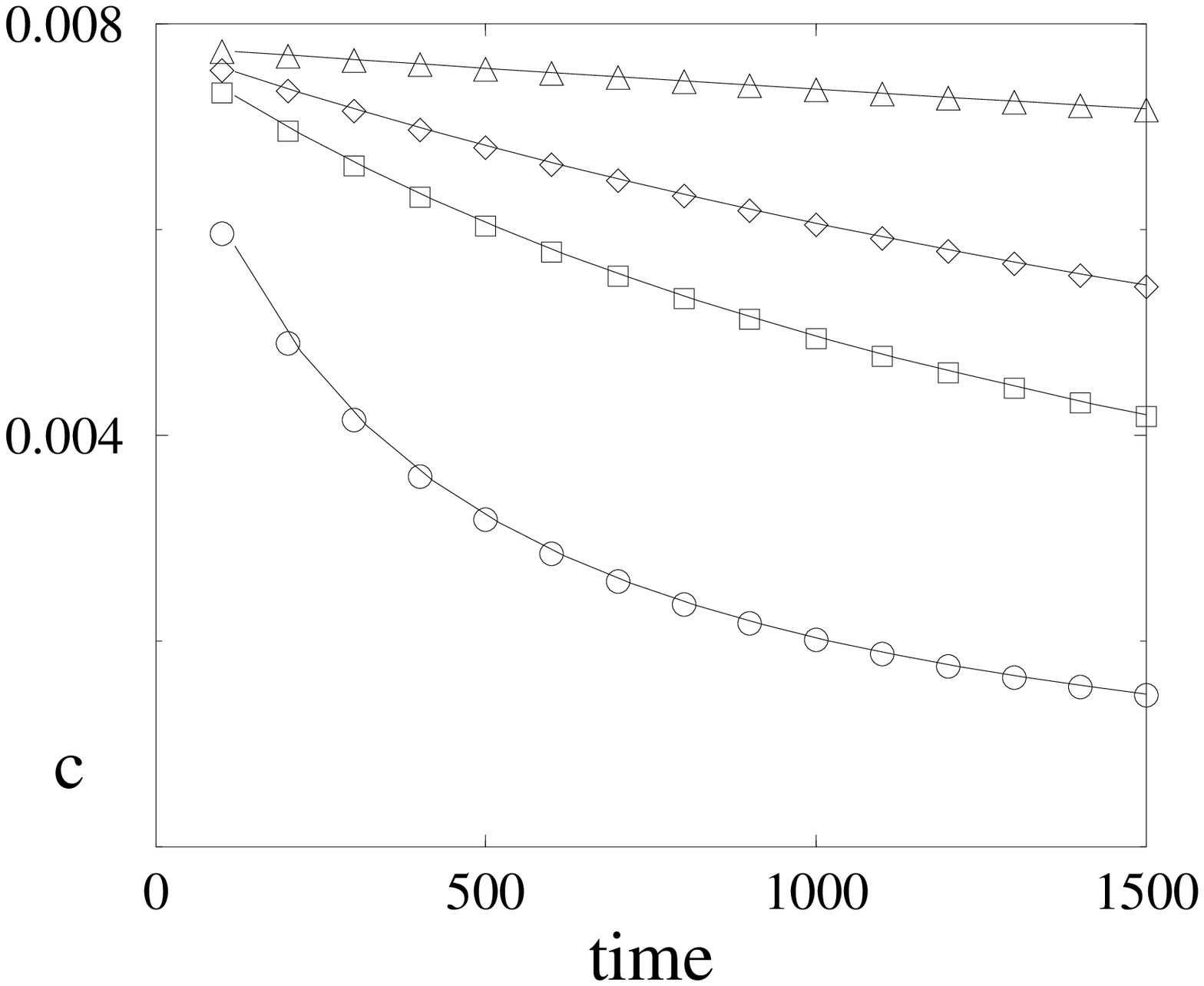,height=6cm,angle=0.,clip=}
\end{center}
\caption {(a) Linear growth rate $Re[\sigma]$ and (b) phase velocity $c$ vs.
time for single mode solutions with $\alpha = \alpha_0 \exp(\epsilon H)$ and
$\Delta = \Delta_0\exp(\epsilon H)$.  Parameter values are $k = 0.85$,
$\alpha_0 = 0.006$, $\Delta_0 = 0.01$, $\Gamma = 0.1$, $\epsilon = 0.5$
circle, $0.1$ square, $0.05$ diamond, $0.01$ triangle.  Continuous lines are
from numerical solution of the full equations, symbols from the approximate
linear stability analysis.}
\label{fig6}
\end{figure}

To investigate the possibility of generating corrugation patterns, we
examine the simplest nontrivial case, namely where $\Delta$ and $\alpha$
have the same $H$-dependence, $\Delta(H) = \Delta_0 \exp(\epsilon H)$ and
$\alpha(H) = \alpha_0 \exp(\epsilon H)$.  As above, this ansatz leads to
$\beta/\alpha$ being constant -- specifically, equal to $\epsilon$ -- and
all three parameters $\alpha$, $\beta$, and $\Delta$ decreasing as $t^{-1}$
at long times $t$.  From (\ref{smallparms}) we then see that the linear
growth rate $\sigma$ is also proportional to $t^{-1}$ for large $t$.
Figures \ref{fig6}a and \ref{fig6}b show this slow decrease of the real and
imaginary parts of $\sigma(t)$ for a few single mode solutions, arrived at
by numerical solution and stability analysis.

Since $\sigma$ is the logarithmic time derivative of the amplitude of a
perturbation, we see that the amplitude itself grows or decays
{\it algebraically\/} in the long-time limit:
\begin{equation}
 \frac{d \ln A}{dt} = \sigma = \frac{\sigma_0}{t+t_0} \qquad \rightarrow
 \qquad A \propto (t+t_0)^{\sigma_0}.
\end{equation}
Thus the growth rate of the corrugation amplitude becomes small at long
times in this model, but the corrugation does not reach a true steady
state.
\begin{figure}[tb]
\begin{center}
\epsfig{file=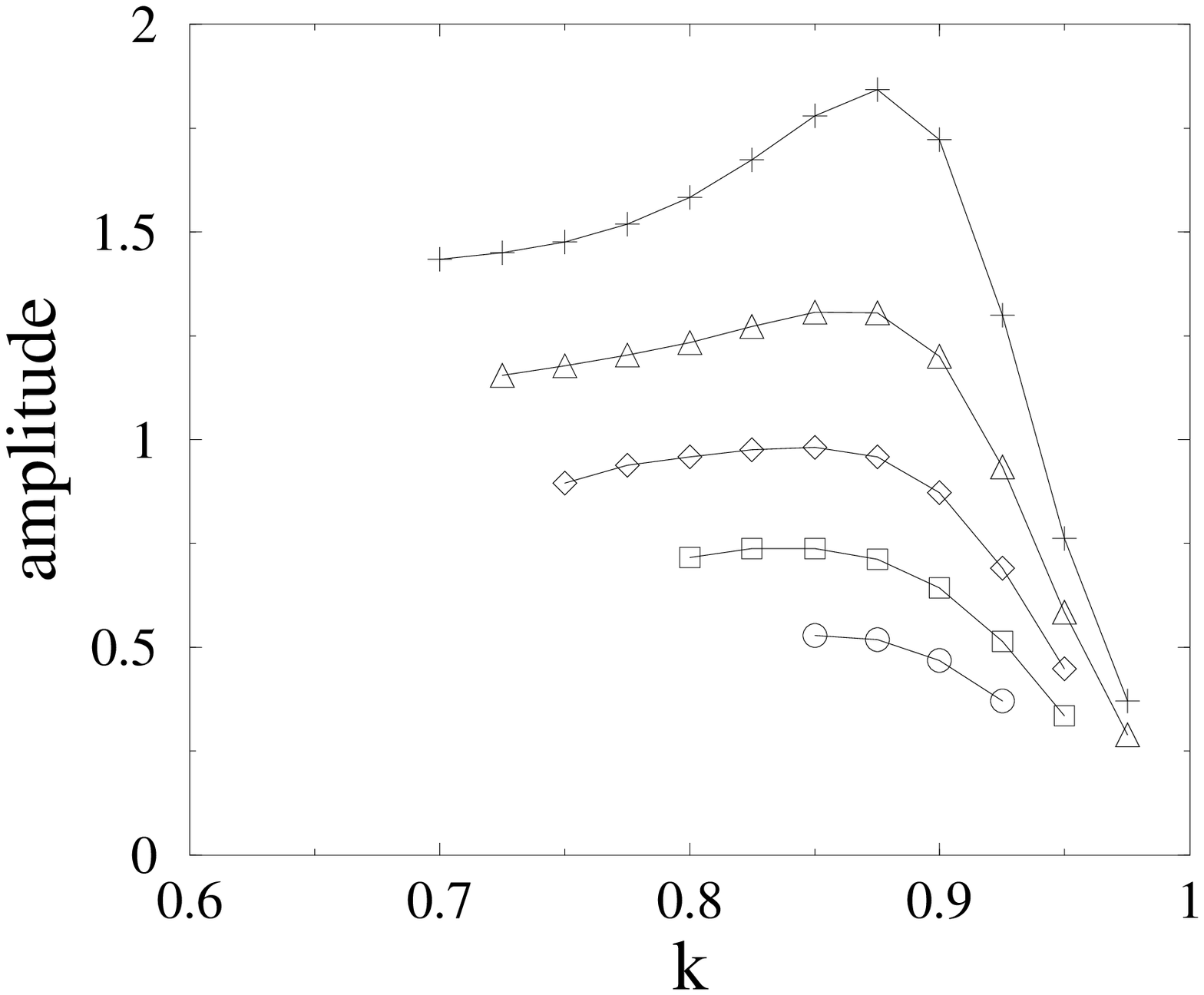,height=6cm,angle=0.,clip=}
\end{center}
\caption{ Quasi-steady state amplitude vs. $k$ in the case of constant
$\alpha$, but setting $\alpha = 0$ when $1+H-Z < 0$.  Parameter values are
$\Gamma = 0.10$, $\Delta = 0.01$, $\alpha = 0.0040$ circle, $0.0045$ square,
$0.0050$ diamond, $0.0055$ triangle, $0.0060$ plus.}
\label{fig7}
\end{figure}

\begin{figure}[tb]
\begin{center}
\epsfig{file=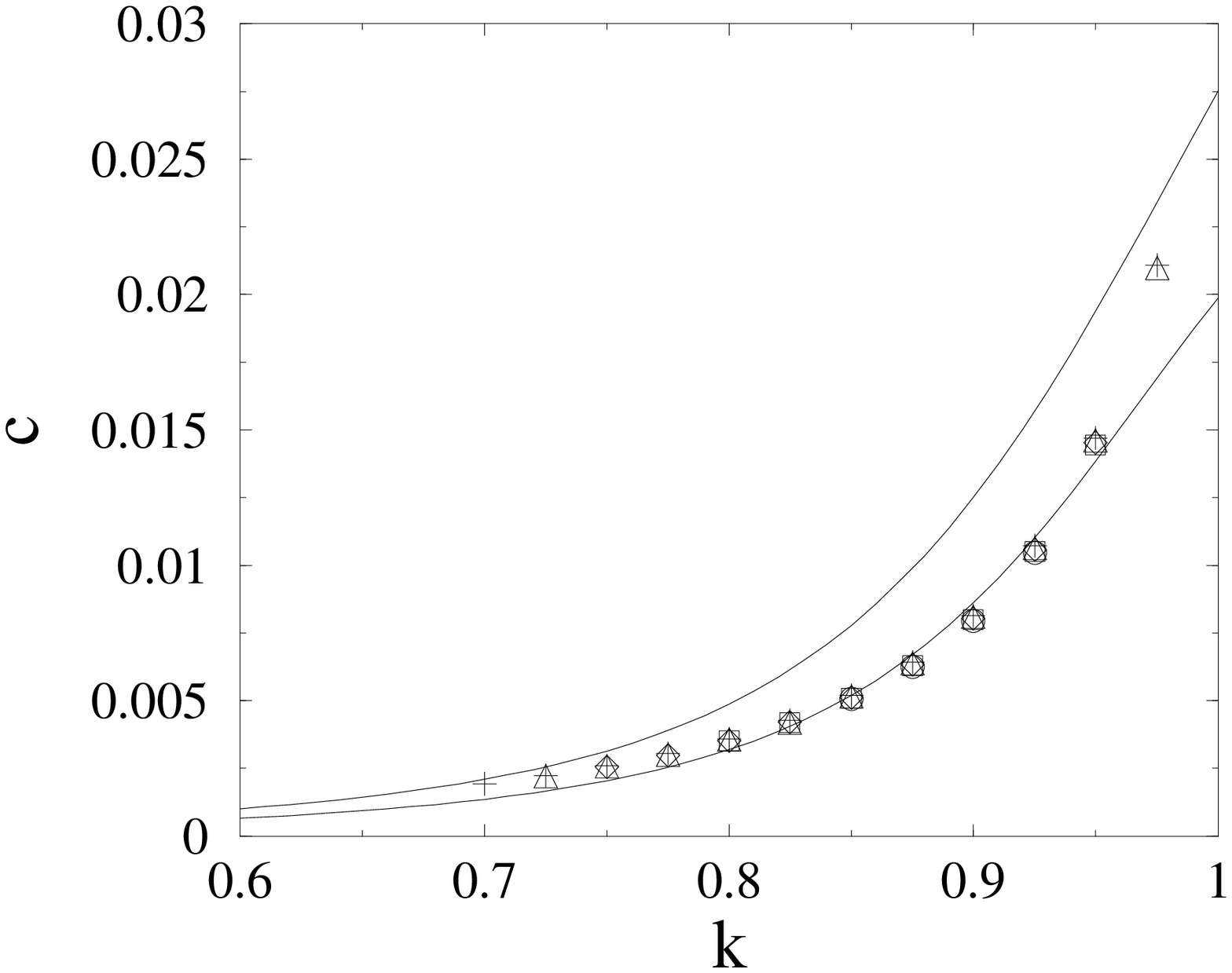,height=6cm,angle=0.,clip=}
\epsfig{file=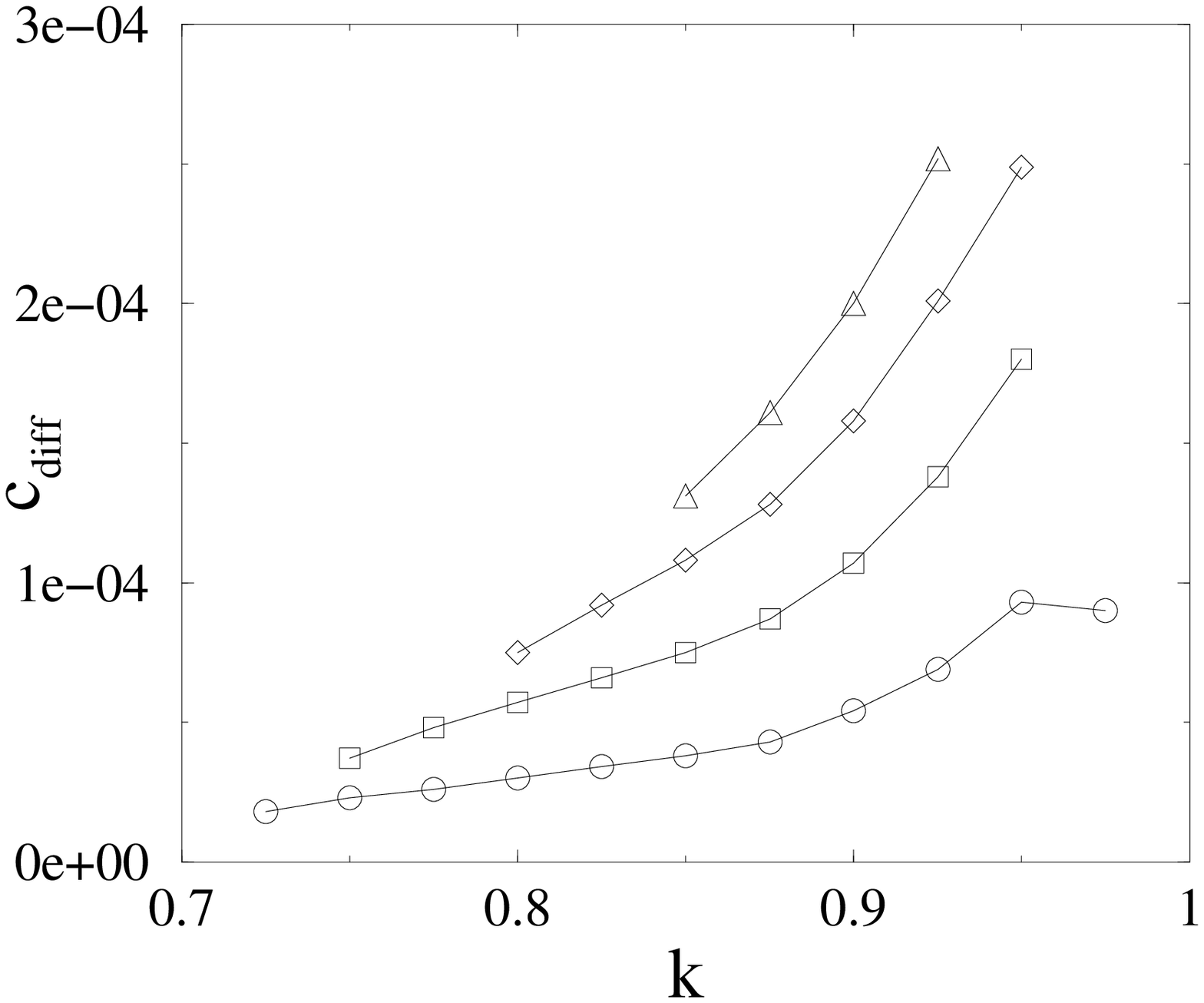,height=6cm,angle=0.,clip=}
\end{center}
\caption{ (a) Quasi-steady state
phase velocity (symbols) $c$ vs. $k$ in the cases described in
Fig. \ref{fig7}.  The
upper and lower solid curves mark where $c(k)$ would lie if the force were
not removed when $1+H-Z < 0$, for $\alpha = 0.0055$ and 0.0040 respectively.
(b) The finer structure in the data presented in (a) is visible here.  We
plot $c_{diff} = c(\alpha = 0.0060) - c(\alpha)$ vs. $k$.  Parameters are
$\alpha = 0.0055$ circle, $0.0050$ square, $0.0045$ diamond, $0.0040$
triangle.}
\label{fig8}
\end{figure}

It is clearly of interest to determine whether restabilized steady states
exist.  However, such states would depend on a host of nonlinear effects
which are omitted in the model embodied in Eqns. (\ref{bigzeqn}) and
(\ref{bigheqn}).  While we do not believe our model as presented is
capable of predicting steady states, we have observed a remarkable feature
in some of our numerical calculations.  We have investigated cases in which
$\alpha$ and $\Delta$ are constants.  These cases neglect the hardening
effect which, as we saw above, tends to stabilize the flat road surface; the
parameter $\beta$ vanishes.  The linear stability analysis is then exact,
and we expect to find purely exponential growth or decay of $h$ and $z$.
Our numerical calculations show that modes that are stable according to the
linear stability analysis do indeed decay in amplitude.  Modes which are
linearly unstable grow until their amplitudes are so large that we have
$1 + H(x,t) - Z(x,t) < 0$ for part of the cycle.  When this happens,
(\ref{bigheqn}) suggests -- unphysically -- that the compression of the road
is negative.  What is happening here is that the cars are losing contact
with the road.  To avoid having the road surface spontaneously rise when an
airborne car passes over it, we set $\alpha = 0$ whenever we are in this
situation.

Such a detachment in fact does occur in real situations, and has been termed
``bouncing'' \cite{CJ71}.  In a laboratory experiment involving two rotating
disks that are in contact with each other under static compression, the two
disks lose contact and bounce against each other when the amplitude of the
corrugation along the perimeter of the disks exceeds about 1/3 the static
compression.  In the case of a car moving on a corrugated surface, such
bouncing will kick the car into the air, but the car will quickly land in a
different place, a mechanism suggested by Mather \cite{Mather}.  Such a
bouncing motion should involve a local flux of the cars and a saltation
function that relates the landing position to the start-off position, as in
the case of wind-blown sand \cite{Bagnold,PT90,NO93}.  We note that such a
nonlocal behavior would be difficult to account for in the model completely,
but our mechanism of setting $\alpha$ locally to zero captures some of its
flavor.  Our expectations are that the modes that grow large enough to be
subject to the local removal of the the force term in (\ref{bigheqn}) will
obviously have their growth rates reduced from those predicted by the linear
stability analysis in the absence of such a force removal, and that this
reduction in growth rate may be sufficient to generate steady states with
finite amplitudes.  To test this intuition, we numerically solve the system
using single cycles of sinusoidal modes $k$, in periodic systems whose
lengths are a single wavelength.  The initial disturbances are chosen with
amplitudes large enough to cause $1 + H(x,t) - Z(x,t)$ to be negative on
occasion.

Provided a given mode is unstable according to the linear stability
analysis, our numerical analysis indicates that it evolves toward a
quasi-steady state in which the amplitude and phase velocity $c$ of the road
corrugations remain fixed, even while the average height of the road bed
continues to sink.  Fig. \ref{fig7} shows the steady state amplitudes of
$h$ as a function of $k$ for $\Gamma$, $\Delta$, and$\alpha$ as given in the
figure caption.  For fixed $k$, the steady state amplitudes increase with
increasing $\alpha$.  Thus we see for this choice of parameters that softer
roads (i.e., larger $\alpha$) support larger disturbances in the steady
state than do harder ones.  This result is intuitively satisfying, as we
expect the ease of deformation of the material to correlate with the
amplitude of the disturbance it supports.  The phase velocities of the
quasi-steady states exhibit particularly interesting behavior.  Fig.
\ref{fig8}a shows the observed quasi-steady state phase velocities, plotted
as symbols, and an ``envelope'' in the $k-c$ plane.   If there were no step
function in the term in $\alpha$, the phase velocities would lie in the
region bounded by the envelope.  The top curve would be the velocity curve
for the largest $\alpha$ used (0.0055) and the bottom curve would be the
velocity curve for the smallest $\alpha$ used (0.0040).  The step function
in the compactivity has forced the phase velocities to ``collapse'' so that
they seem to lie along or near a single curve in the $k-c$ plane.
Examination of the quasi-steady state velocities on a finer scale, as shown
in Fig. \ref{fig8}b, indicates a persistence of the kind of variation of
phase velocity with softness we have typically seen (that is, the softer the
road, or equivalently the larger $\alpha$ is, the greater the phase
velocity, all other things being equal), though this variation exists on a
much finer scale in the steady state case than in the case in which the step
function is not invoked.

Even though we have identified interesting steady state behavior in the
solution of the differential equations in this limiting case of $\alpha$
being constant, we must reiterate that setting $\alpha$ to a constant is
certainly unphysical, and that many nonlinear effects have been omitted
from our model equations.  Moreover, our model does not account carefully
for what happens when contact with the road and cars is lost; it merely
turns off the coupling between cars and road in the equation of motion for
$H(x,t)$, and furthermore assumes the evolution of $Z(x,t)$ remains well
described by Eq. (\ref{bigzeqn}).  Thus we cannot expect that our nonlinear
calculations will speak to the behavior of real roads.  Nonetheless, we
find it remarkable that simply setting $\alpha = 0$ when
$1 + H(x,t) - Z(x,t)$ is negative seems to contain the growth of the
unstable modes.

\section{Conclusions and Prospects}

In this work, we have explored the origin of the corrugation instability of
dirt roads subjected to a constant flux of cars.  We have presented a simple
phenomenological model for the evolution of the road surface and carried out
a linear stability analysis to uncover the gross features of the
instability.
>From our approximate stability analysis and its numerical verification, we
find that the dynamical processes of diffusion and compression in a road bed
are sufficient to generate instabilities that can give rise to pattern
formation.  Small-amplitude corrugations on the road surface grow when the
diffusion parameter $\Delta/\alpha$ and the hardening parameter
$\beta/\alpha$ lie below the stability boundary (for the appropriate damping
coefficient $\Gamma$ for the cars) shown in Figure \ref{fig3}.  From the
definitions (\ref{parmdefs}) of $\alpha$ and $\Delta$ in terms of the
original, dimensional softness and diffusion coefficients $a(H)$ and $D(H)$,
we see that the dimensionless combination $\Delta/\alpha$ is inversely
proportional to the square of $v_x$, the horizontal speed of the cars across
the road.  Thus we find that there is a critical car speed above which the
flat road surface is unstable.  Also, since $a(H)$ is proportional to the
flux of cars, there is a critical flux for any given speed, above which the
flat road is unstable.  Both diffusive relaxation and hardening of the
surface are stabilizing effects.  The wave number of the most rapidly
growing mode is given implicitly in dimensionless terms by (\ref{kmaxeqn}),
which is plotted in Fig. \ref{fig4}.

Our model is clearly schematic; a quantitative theory of road corrugations
would need to incorporate a number of effects we left out of our
calculations.  We have not attempted to account, for instance, for the
geometric distinction between $\partial h/\partial t$ and the normal
velocity
of the road surface as it compacts, nor for the fact that the force exerted
by a moving car on a corrugated road surface is not purely vertical.  For
these reasons alone we have not carried out any nonlinear analysis on our
model -- the model is explicitly not valid beyond linear order in the
amplitude of the incipient corrugation.  We have also ignored all details of
the physical processes by which the road compacts, so it is not clear how
well our phenomenological picture of compaction being proportional to
applied
vertical force or impulse can describe the dynamics of a real road.  Our
model for the cars is the simplest possible, neglecting the multiplicity of
oscillation modes of real cars and the Hertzian nature of the contact force
between the tires and the road.  In particular, the fact that real vehicles
have two or more pair of tires separated by fixed distances may introduce
a new, relevant length scale into the problem.  Finally, a quantitative
theory would have to incorporate distributions of vehicle sizes, weights,
oscillation frequencies, speeds, and the like.

Despite the fact that most nonlinear effects are omitted from our model,
we find particularly interesting the numerical result that simply setting
the contact force to zero when the equations would make it negative can lead
to an apparently steady state.  At least this results in a drastic slowing
of the dynamics once a certain road configuration has been established.
Continuing work will focus on determining the selection mechanism of such
``frozen'' states, and their response to local perturbations.


\begin{thebibliography}{99}
\bibitem{SSC82} J. Stoddart, R. Smith, and R. M. Carson, Transp. En. J.
{\bf 108}, 376 (1982).

\bibitem{Mather} K. B. Mather, Civ. Eng. Public Works Rev. {\bf 57}, 617,
781 (1962); Scientific American {\bf 206}(1), 128 (1963).

\bibitem{Papad} J. Papadopoulos, private communication.

\bibitem{Relton} F. E Relton, Roads and Road Constr. {\bf 16}, 340 (1938).

\bibitem{Bagnold} R. A. Bagnold, Proc. Roy. Soc. A157, 594 (1936);
See also {\it The Physics of Blown Sand and Desert Dunes\/}
(Morrow, New York, 1941).

\bibitem{PT90} K. Pye and H. Tsoar, {\it Aeolian Sand and Sand Dunes\/}
(Unwin Hyman, London, 1990).

\bibitem{NO93} H. Nishimori and N. Ouchi, Phys. Rev. Lett.
{\bf 71}, 197 (1993).

\bibitem{KBH00} D. A. Kurtze, J. A. Both, D. C. Hong, Phys. Rev. E
{\bf 61}, 6750 (2000).

\bibitem{CJ71} R. M. Carson and K. L. Johnson, Wear, {\bf 17}, 59 (1971).

\end{thebibliography}
\end{document}